\documentclass[aps,pre,singlecolumn,superscriptaddress]{revtex4}

\usepackage{amsmath, amsfonts}
\usepackage{graphicx} 
\usepackage{physics}  
\usepackage{color}    
\usepackage{tikz}     
\usepackage{subcaption} 
\usepackage{float}    
\usepackage{mathtools} 
\usepackage[english]{babel} 
\usetikzlibrary{arrows.meta, calc, shapes.misc}

\newcommand{\beq}{\begin{equation}}
\newcommand{\eeq}{\end{equation}}
\newcommand{\eps}{\varepsilon}

\begin{document}

\title{Exponential Asymptotics for Translational Modes in the Discrete Nonlinear Schr{\"o}dinger Model}

\author{C. J. Lustri}
\affiliation{School of Mathematics and Statistics, The University of Sydney, Camperdown 2050, Australia}

\author{P. G. Kevrekidis}
\affiliation{Department of Mathematics and Statistics, University of Massachusetts Amherst, Amherst, MA 01003, USA}

\affiliation{Department of Physics, University of Massachusetts Amherst, Amherst, MA 01003, USA}

\author{S. J. Chapman}
\affiliation{Mathematical Institute, University of Oxford, Radcliffe Observatory, Andrew Wiles Building, Woodstock Rd, Oxford OX2 6GG, UK}

\begin{abstract}
In the present work, we revisit the topic of translational eigenmodes in discrete models. We focus on the prototypical example of the discrete nonlinear Schr{\"o}dinger equation, although the methodology presented is quite general. 
We tackle the relevant discrete system based on exponential asymptotics and start by deducing the well-known (and fairly generic) feature of the existence of two types of fixed points, namely site-centered and inter-site-centered. Then, turning to the stability problem, we not only retrieve the exponential scaling (as \( e^{-\pi^2/(2 \varepsilon)} \), where \( \varepsilon \) denotes the spacing between nodes) and its corresponding prefactor power-law (as \( \varepsilon^{-5/2} \)), both of which had been previously obtained, but we also obtain a highly accurate leading-order prefactor and, importantly, the next-order correction, for the first time, to the best of our knowledge. 
This methodology paves the way for such an analysis in a wide range of lattice nonlinear dynamical equation models.
\end{abstract}

\maketitle

\section{Introduction}

It is well-known (and has by now been extensively studied) that
lattice nonlinear dynamical equations of dispersive type arise
in a wide range of applications. These emerge, among others,
in the dynamics of the electric field of light in optical
waveguides~\cite{LEDERER20081}, as well as in the mean-field
dynamics of atomic Bose-Einstein
condensates (BECs) in deep optical lattices~\cite{RevModPhys.78.179}.
They are also relevant to nonlinear coupled lattices of electrical
circuits governed by Kirchhoff's laws applied to nonlinear elements
(such as capacitors)~\cite{remoissenet}, as well as to coherent
structures arising in elastic media and mechanical metamaterials
such as granular crystals and variants thereof~\cite{Nester2001,yuli_book}.
Further settings of considerable impact involve the 
arrays of superconducting Josephson junctions in condensed
matter physics~\cite{alex,alex2}, as well as the denaturation
models of the DNA double strand~\cite{peybi,Michel_Peyrard_2004}.

Admittedly, there exists a wide variety of associated models,
with different ones among them being of interest/relevance
to different applications. For instance, the quintessential
---within nonlinear science--- models of the 
Fermi-Pasta-Ulam-Tsingou type~\cite{FPUreview,VAINCHTEIN2022133252} are particularly
well suited to granular systems, while discrete Klein-Gordon
(variants of sine-Gordon) lattices~\cite{imamat} are tailored to superconducting 
Josephson junctions. However, arguably, a central
envelope model that emerges in a wide range of applications
consists of the discrete nonlinear Schr{\"o}dinger (DNLS) 
equation~\cite{kev09,chriseil}.
For the latter, the most canonical framework consists
of arrays of optical waveguides (where its relevance 
and extensive findings, including in experiments, have been
summarized in various reviews~\cite{LEDERER20081,cole}), but 
it has also been of value to
other fields such as BECs with optical lattice potential wells
forming the effective lattice nodes~\cite{RevModPhys.78.179,Alfimov2}.

In all of these models bearing on-site potentials (discrete Klein-Gordon (DKG)
and DNLS ones, for instance), there exist some principal features
that clearly distinguish the discrete case from the continuum one.
Arguably, the most notable one is the breaking of a fundamental
continuum symmetry of the problem, the translational invariance
thereof. The latter is well-known through Noether theory to be 
associated with the conservation of momentum~\cite{noether1971invariant}.
This, in turn, has significant consequences as concerns the prototypical
solitary waves of the models~\cite{carretero2025nonlinear}. Indeed,
typically it is found that such models, instead of a single discrete
analogue to a continuum wave, they bear two such discrete analogues.
One of them is referred to as site-centered and the other as bond-centered
or inter-site-centered. Perhaps even more importantly, one of these
states is found to be spectrally stable (typically in DKG and DNLS
models it is the site-centered one) and one is spectrally unstable
(typically, the inter-site-centered one). This is because the
invariance of translation within the continuum is reflected in a 
zero eigenvalue (pair --- as the system is Hamiltonian and eigenvalues
come in pairs), which upon the perturbation of discreteness
can either move along the imaginary axis, in the spectrally stable
site-centered case, or along the real axis, in the spectrally unstable
bond-centered alternative.

It was realized a considerable while ago~\cite{Todd_Kapitula_2001} (see also references therein)
that the relevant bifurcation of this translational eigenmode
has to be exponentially small. A heuristic argument to that effect
hinges on the Taylor expansion of the second difference 
(see also below), which to all algebraic orders in the relevant
lattice spacing parameter can be approximated by a sum involving
even derivatives, none of which breaks translational invariance.
Hence, the bifurcation of the relevant eigenvalue from the
origin of the spectral plane of eigenvalues {\it needs} to be
exponentially small. Indeed, the work of~\cite{Todd_Kapitula_2001}
sought to approximate this eigenvalue that was
also observed in numerous other works, including, e.g.,
in~\cite{johaub,kev09} etc. However, the relevant work
developed a perturbative argument from the integrable
analogue of the equation ---the so-called Ablowitz-Ladik
model~\cite{AblowitzPrinariTrubatch}. This had, however,
a distinct drawback: while it seemed to capture the
correct exponentially small term (and even the prefactor
thereof as concerns its scaling power on the lattice spacing), 
it was argued therein that the prefactor could not be
properly captured, nor could corrections to this leading
order behavior be assessed.

In the present work we  definitively address the
relevant problem. Indeed, we utilize an exponential asymptotics
technique based on \cite{Chapman, Daalhuis, king_chapman_2001} (partially reminiscent of what was done
for continuum problems with periodic potentials
in the work of~\cite{HWANG20111055}) that enables us
to identify the two different possible stationary solitary
waves of the lattice. We then bring the corresponding result
to the spectral stability analysis problem and observe
its implications on the (formerly zero translational) eigenvalue. 
Our technique allows us to indeed compute definitively
the relevant prefactor and even lends itself to the calculation
of the next order correction. By using the rather numerically
stringent test of multiplying the eigenvalue by the relevant
exponential (and leveraging an accurate computation of the
relevant eigenvalue down to the limit of double precision),
we are able to showcase the asymptotic agreement of our
prediction with the numerical findings. What's more,
this methodology provides the reader with a tool that 
is expected to naturally generalize to other models carrying
this generic ---within lattice dynamical settings--- feature.

Our presentation is structured as follows. In section II, we briefly
provide the model's mathematical setup and notation. In section III,
we focus on the existence problem and its corresponding solitary wave
solution. Section IV details our stability findings. 
Finally, in section V, we summarize our results and present
some possibilities for future studies.

\section{Model Setup}

The 1D discretized nonlinear Schr\"{o}dinger (DNLS) equation is given 
by~\cite{kev09}
\begin{equation}
{\rm i} \frac{dQ_n}{dt} + \Delta_2 Q_n + |Q_n|^2 Q_n = 0,\label{e:Qn}
\end{equation}
where the discrete Laplacian associated with the spacing 
$\varepsilon$ reads:
\begin{equation}
\Delta_2 Q_n = \frac{Q_{n-1} - 2 Q_n + Q_{n+1}}{\eps^2}.\label{e:Delta2}
\end{equation}

The continuous 1D nonlinear Schr\"{o}dinger (NLS) equation, i.e., the
continuum limit of the model as $\varepsilon \rightarrow 0$ assumes the form:
\begin{equation}
{\rm i}q_t + q_{xx} + |q|^2 q = 0.\label{e:NLS}
\end{equation}

While we will not tackle them in the present work, it is relevant
to point out that our considerations can be extended
firstly to the setting of
the 1D discretized nonlinear Schr\"{o}dinger equation involving next-to-nearest neighbors: 
\begin{equation}
\Delta_2 Q_n = -\frac{1}{12}Q_{n-2} + \frac{4}{3}Q_{n-1} - \frac{5}{2}Q_n + \frac{4}{3}Q_{n+1} - \frac{1}{12}Q_{n+2}.
\end{equation}
and secondly to the more general realm of next-nearest neighbor 
interactions (cooperating or potentially competing with the nearest
neighbor ones) in the modeling form:
\begin{equation}
\mu(Q_{n+2}-2Q_n+Q_{n-2})+ {\rm i} \frac{dQ_n}{dt} + \Delta_2 Q_n + |Q_n|^2 Q_n = 0
\end{equation}
which is inspired by setups
in the form of zigzag waveguide arrays~\cite{Efremidis2002,Szameit:09}.

\section{Series Solution}
Let us use the well-known standing wave
form $Q_n = {\rm e}^{{\rm i} \omega t} q_n$ in \eqref{e:Qn} to obtain the stationary DNLS,
\begin{equation}
-\omega q_n + \frac{1}{\varepsilon^2}({q_{n+1}-2q_n + q_{n-1}})+ |q_n|^2 q_n = 0.
\end{equation}
We take a continuum limit by setting $z= \eps n$ and letting $q_n \equiv q(z)$ to obtain
\begin{equation}
-\omega q(z) + \frac{1}{\eps^2}\left(q(z+\eps) - 2q(z) + q(z-\eps)\right) + |q(z)|^2 q(z) = 0.
\end{equation}
Expanding this as a power series in $\eps$ gives the infinite-order differential equation
\begin{equation}\label{e:infode}
-\omega q(z) + 2\sum_{m=1}^{\infty} \frac{\varepsilon^{2m-2}}{(2m)!}\frac{d^{2j}q(z)}{dz^{2j}} + |q(z)|^2 q(z) = 0.
\end{equation}
We expand $q(z)$ as an asymptotic power series in $\eps$ such that
\begin{equation}\label{e:series}
q(z) \sim \sum_{j=0}^{\infty} \eps^{2j}q_j(z) \quad \mathrm{as} \quad \eps \to 0.
\end{equation}
The leading-order solution $q_0(z)$ satisfies the (ODE resulting from the standing wave ansatz within the) continuous nonlinear Schr\"{o}dinger equation \eqref{e:NLS}. We select the soliton solution, given by
\begin{equation}\label{e:q0}
q_0(z) = \sqrt{2\omega}\sech[\sqrt{\omega}(z-z_0)].
\end{equation}
Notice that here without loss of generality, and taking advantage
of the phase/gauge invariance of the model~\cite{kev09,carretero2025nonlinear}, we have selected
the real solution to the relevant problem.
For simplicity in the subsequent analysis, we define $\tilde{z} = z-z_0$, and write our equations in terms of this shifted variable. Substituting \eqref{e:series} into \eqref{e:infode} and noting that $q_0$ is real and positive gives the recurrence relation
\begin{equation}\label{e:recur}
-\omega q_j + 2\sum_{m=1}^{\infty} \frac{1}{(2m)!} \frac{d^{2m} q_{j-m+1} }{d\tilde{z}^{2m}}+ 2 q_0^2 |q_j |+ q_0^2 q_j = 0.
\end{equation}
It will be useful to identify certain asymptotic properties of $q_0$. The tail asymptotics of $q_0$ are given by
\begin{equation}\label{e:tail}
q_0(\tilde{z}) \sim 2\sqrt{2 \omega}e^{-\sqrt{\omega}\tilde{z}}\quad \mathrm{as}\quad \tilde{z} \to \infty.
\end{equation}
The leading-order solution is singular at
\begin{equation}\label{e:zs}
 \tilde{z}_s = \pm\frac{1}{\sqrt{\omega}}\left[\frac{\pi {\rm i}}{2} + k \pi {\rm i} \right],
\end{equation}
where $k \in \mathbb{Z}$. We will denote specific choices of $k$ as $\tilde{z}_{s,k}$ where appropriate. Near these points, we find
\begin{equation}\label{e:q0nearzs}
q_0(\tilde{z}) \sim -\frac{ {\rm i}\sqrt{2}}{\tilde{z}-\tilde{z}_s} + \frac{{\rm i}\omega}3{\sqrt{2}}(\tilde{z}-\tilde{z}_s)\quad \mathrm{as}\quad \tilde{z} \to \tilde{z}_s.
\end{equation}
The dominant exponentially small oscillations will arise from the singularities $\tilde{z}_{s,0}$ and $\tilde{z}_{s,-1}$. By rearranging \eqref{e:recur}, we can see that each term $q_j$ is obtained by repeatedly differentiating previous terms in the series, and we therefore expect singularities in $q_j$ for $j > 0$ to also be located at the points in \eqref{e:zs}. The singularities will become stronger as $j$ increases.

Continuing the calculated terms using the recurrence relation \eqref{e:recur} gives
\begin{equation}
    \frac{d^2 q_1}{d\tilde{z}^2} + \frac{1}{12}\frac{d^4 q_0}{d\tilde{z}^4} + (3q_0^2 - \omega)q_1 = 0.
\end{equation}
Solving this gives the particular solution
\begin{equation}\label{e:q1}
q_1 = \frac{\omega^{3/2}(9-7\cosh(2\sqrt{\omega}\tilde{z}) + \sqrt{\omega}\tilde{z} \sinh(2\sqrt{\omega}\tilde{z}))}{24\sqrt{2}\cosh^3(\sqrt{\omega}\tilde{z})}.
\end{equation}
Near the singularities at $\tilde{z} = \tilde{z}_{s}$, this has the asymptotic behaviour
\begin{equation}\label{e:q1nearzs}
    q_1(\tilde{z}) \sim \frac{{\rm i}\sqrt{2}}{3(\tilde{z}-\tilde{z}_s)^3} + \frac{\pi\sqrt{\omega}}{24\sqrt{2}(\tilde{z}-\tilde{z}_s)^2}\quad \mathrm{as} \quad \tilde{z} \to \tilde{z}_{s}.
\end{equation}

\subsection{Late-Order Asymptotics}

We now need to determine the asymptotic behaviour of $q_j(\tilde{z})$ as $j \to \infty$. In singularly perturbed equations such as \eqref{e:infode}, obtaining $q_j$ requires repeatedly differentiating earlier terms in the series. This repeated differentiation causes the singularity strength to increase at subsequent orders, and the series terms to diverge in a predictable factorial-over-power fashion \cite{Dingle}. Based on this observation, \cite{Chapman} proposed that the asymptotic behaviour of $q_j$ as $j \to \infty$ can be written a sum of factorial-over-power terms with the form
\begin{equation}\label{e:ansatz}
q_j(\tilde{z}) \sim \frac{Q(\tilde{z})\Gamma(2j + \gamma)}{\chi(\tilde{z})^{2j+\gamma}},
\end{equation}
where $\chi(z) = 0$ at some $\tilde{z} = \tilde{z}_s$, and each singular point $\tilde{z}_s$ generates late-order terms in the asymptotic behaviour. 

We will determine $Q(\tilde{z})$, $\chi(\tilde{z})$ and $\gamma$. By substituting this expression into \eqref{e:recur}, we obtain
\begin{align}\nonumber
-\frac{\omega Q\Gamma(2j+\gamma)}{\chi^{2j+\gamma}} + 2\sum_{m=1}^{j-1} \Bigg[\frac{(\chi')^{2m}}{(2m)!}\frac{Q\Gamma(2j + 2 + \gamma)}{\chi^{2j+2+\gamma}} - \frac{(\chi')^{2m-1}}{(2m-1)!}&\frac{Q'\Gamma(2j + 1 + \gamma)}{\chi^{2j+1+\gamma}} \\
+\frac{(\chi')^{2m-2}}{(2m-2)!}\frac{Q''\Gamma(2j  + \gamma)}{\chi^{2j+\gamma}}\Bigg] &+\frac{3q_0^2Q\Gamma(2j+\gamma)}{\chi^{2j+\gamma}} + \ldots = 0,\label{e:LOT.1}
\end{align}
where the terms that we have retained are all $\mathcal{O}(q_j)$ as $j \to \infty$, and all omitted terms are smaller in this limit than those retained, and $q_j(\tilde{z})$ is taken to be real-valued. Note that we omitted any term proportional to $\chi''$, which would appear at order $\mathcal{O}(q_{n+1/2})$, because we will find at $\mathcal{O}(q_{j+1})$ that $\chi'$ is constant, and therefore $\chi'' = 0$. 

\subsubsection{Calculating the singulant, $\chi$}
Balancing terms of $\mathcal{O}(q_{j+1})$ in \eqref{e:LOT.1} as $j \to \infty$ gives
\begin{equation}\label{e:singulanteq}
 2 Q \sum_{m=1}^{j-1} \frac{(\chi')^{2m}}{(2m)!} = 0.
 \end{equation}
We can determine the leading-order solution of this in the limit that $j \to \infty$ by extending the upper bound of the summation term to infinity. This gives $\cosh(\chi') = 1$, or
 \begin{equation}\label{e:chi}
\chi = 2\pi{\rm i}\ell\left(\tilde{z} - \tilde{z}_s\right),
\end{equation}
where $\ell \in \mathbb{Z}$ and the final term corresponds to the positive choice of sign in \eqref{e:zs} with $k = 0$. We select $\ell = 1$ as the solution of interest, as this corresponds to the largest exponentially small term associated with the singularity at $\tilde{z}_{s,0}$.  

\begin{figure}[tb]
\centering
\begin{tikzpicture}
[xscale=1,>=stealth,yscale=1]

\fill[blue!10] (0,-2.8) -- (3.5,-2.8) -- (3.5,2.8) -- (0,2.8) -- cycle;
\draw[->] (-4,0) -- (4,0) node[right] {{$\mathrm{Re}(\tilde{z})$}};
\draw[->] (0,-3.2) -- (0,3.2) node[above] {{$\mathrm{Im}(\tilde{z})$}};
\draw[blue,line width=1mm] (0,-2) -- (0,2);
\draw[blue,line width=1mm,dotted] (0,-2) -- (0,-2.8);
\draw[blue,line width=1mm,dotted] (0,2) -- (0,2.8);
\draw[line width=0.7mm] (-0.15,1.85) -- (0.15,2.15);
\draw[line width=0.7mm] (-0.15,2.15) -- (0.15,1.85);
\draw[line width=0.7mm] (-0.15,-1.85) -- (0.15,-2.15);
\draw[line width=0.7mm] (-0.15,-2.15) -- (0.15,-1.85);
\node[left] at (-0.275,2) {$\dfrac{{\rm i}\pi}{2\sqrt{\omega}}$} ;
\node[left] at (-0.275,-2) {$-\dfrac{{\rm i}\pi}{2\sqrt{\omega}}$} ;

\node at (-2,-0.5) {{$q_{\mathrm{exp}} = 0$}};
\node at (2,-0.5) {{$q_{\mathrm{exp}} \neq 0$}};

\fill[blue!10] (5.5,2.3) -- (5.5,2.8) -- (6,2.8) -- (6,2.3) -- cycle;
\draw (5.5,2.3) -- (5.5,2.8) -- (6,2.8) -- (6,2.3) -- cycle;
\draw (5.5,1.3) -- (5.5,1.8) -- (6,1.8) -- (6,1.3) -- cycle;
\draw[line width=0.7mm] (5.6,0.7) -- (5.9,0.4);
\draw[line width=0.7mm] (5.6,0.4) -- (5.9,0.7);
\draw[blue,line width=1mm] (5.75,-0.2) -- (5.75,-0.7);
\draw[blue,line width=1mm,dotted] (5.75,-1.2) -- (5.75,-1.7);
\node[right] at (6.1,2.55) {Exponentially small contribution};
\node[right] at (6.1,1.55) {No exponentially small contribution};
\node[right] at (6.1,0.55) {Leading-order singularity};
\node[right] at (6.1,-0.45) {Primary Stokes line};
\node[right] at (6.1,-1.45) {Additional Stokes lines};

\draw[gray] (3.5,2.8) -- (3.5,-2.8) -- (-3.5,-2.8) -- (-3.5,2.8) -- cycle;
\end{tikzpicture}
\caption{Stokes structure for the discrete 
solitary wave solution. We could pick the exponentially small term to be zero on either the left or the right-hand side of the Stokes line, shown as a solid blue line; this choice does not affect the final result. There are additional singularities further along the imaginary axis in both directions that produce additional Stokes lines on the line $\mathrm{Re}(\tilde{z}) = 0$, represented as dotted blue lines, but these Stokes lines switch on asymptotic contributions that are exponentially small compared to those considered here.}\label{fig:stokes}
\end{figure}
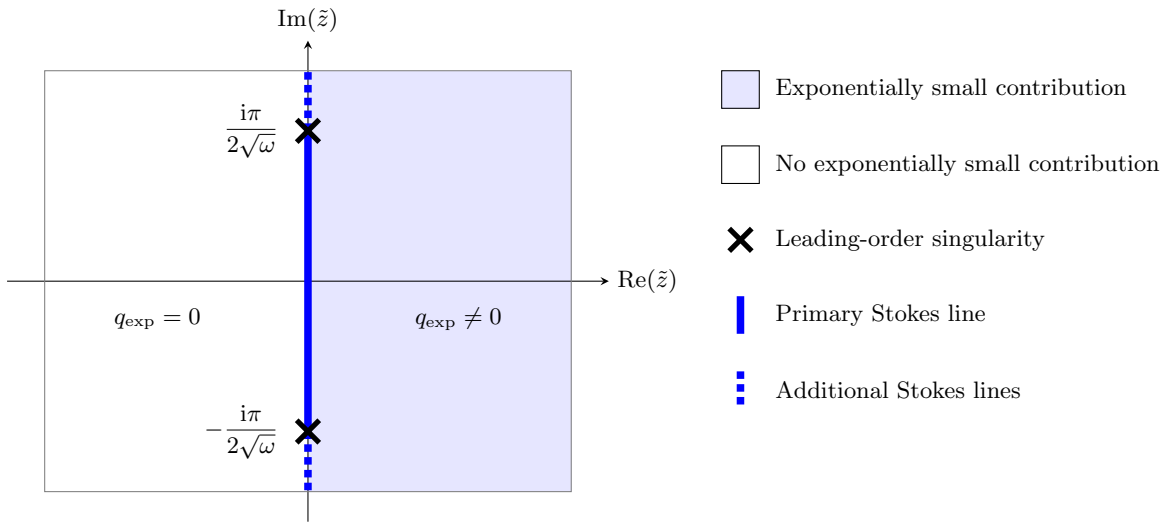

From this, we conclude that there is a Stokes line in the solution, along the curve satisfying
\begin{equation}
    \mathrm{Im}(\chi) = 0,\qquad \mathrm{Re}(\chi) > 0. 
\end{equation}
This configuration is shown in Figure \ref{fig:stokes}, with the Stokes line being represented as a solid blue line. Without loss of generality, we will select the contribution so that it is zero on the left-hand side of the Stokes line, and is switched on as the line is crossed from left ($\tilde{z}<0$) to right ($\tilde{z}>0$). 

From \eqref{e:chi}, we see that $\chi^*$, or the complex conjugate of $\chi$ is a solution to this equation, and this corresponds to the largest exponentially small term associated with the singularity at $\tilde{z}_{s,-1}$. This will produce a second Stokes line along the curve $\mathrm{Re}(\tilde{z}) = 0$, across which another exponentially small contribution appears. This contribution will be the complex conjugate of the contribution associated with $\chi$ on the real axis, meaning that the total exponentially small contribution is real for $\tilde{z} \in \mathbb{R}$.

\subsubsection{Calculating the prefactor, $Q$}
At $\mathcal{O}(q_{j+1/2})$ as $j \to \infty$, we balance terms in \eqref{e:LOT.1} to obtain
\begin{equation}
 2\sum_{m=1}^{j-1} \frac{Q'(\chi')^{2m-1}}{(2m-1)!} = 0.
 \end{equation}
As $\chi' = 2\pi{\rm i}\ell$, this statement is always true when $j \to \infty$. This is a generic feature of such discrete systems, and appears in any discrete equation with only second-order differences. 

Balancing terms at the next order, $\mathcal{O}(q_j)$, gives
 \begin{align}
2\sum_{m=1}^{j-1} \frac{Q''(\chi')^{2m-2}}{(2m-2)!} + (3 q_0^2  - \omega )Q  &= 0.\label{e:UV.1}
 \end{align}
Extending the summation term to infinity gives, after some algebraic manipulation, 
\begin{equation}\label{e:Uode}
Q'' + (6  \sech^2(\sqrt{\omega}\tilde{z}) - 1) \omega Q  = 0.
\end{equation}

This can be solved to give the prefactor
\begin{equation}
Q(\tilde{z}) = K_1 Q_1(\tilde{z}) + K_2 Q_2(\tilde{z}),\label{e:UK}
\end{equation}
where $K_1$ and $K_2$ are constants that have yet to be determined, and
\begin{align}
Q_1(\tilde{z}) &= \tfrac{1}{2} \tanh(\sqrt{\omega}\tilde{z})\sech(\sqrt{\omega}\tilde{z}),\label{e:U1}\\
Q_2(\tilde{z}) &= \sech(\sqrt{\omega}\tilde{z})\,\left(6\sqrt{\omega}\tilde{z}\tanh(\sqrt{\omega}\tilde{z}) + \cosh(2\sqrt{\omega}\tilde{z})-5\right).\label{e:U2}
\end{align}
 The limiting behaviour as $\tilde{z} \to \infty$ of these solutions is
\begin{equation}\label{e:Qlimit}
Q_1(\tilde{z}) \sim e^{-\sqrt{\omega}\tilde{z}}, \qquad Q_2(\tilde{z}) \sim e^{\sqrt{\omega}\tilde{z}}.
\end{equation} 
The asymptotic behaviour of these solutions as $\tilde{z} \to \tilde{z}_s$ is, 
\begin{align}
Q_1(\tilde{z}) \sim -\frac{\rm i}{2\omega(\tilde{z} - \tilde{z}_s)^2},\qquad Q_2(z) \sim \frac{3\pi}{\omega(\tilde{z}-\tilde{z}_s)^2}.\label{e:Uasymp}
\end{align}

\subsubsection{Calculating $\gamma$}

We use \eqref{e:q0nearzs} in \eqref{e:Uode} to obtain the local governing equation near $\tilde{z} = \tilde{z}_s$,
\begin{equation}
Q'' \sim \frac{6Q}{(\tilde{z}-\tilde{z}_s)^2},
\end{equation}
which tells us that the asymptotic behaviour of the solution is
\begin{equation}\label{e:localU}
Q(\tilde{z}) \sim \Lambda_1 (\tilde{z}-\tilde{z}_s)^3 + \frac{\Lambda_2}{(\tilde{z}-\tilde{z}_s)^2},
\end{equation}
where $\Lambda_1$ and $\Lambda_2$ are arbitrary constants yet to be determined. The asymptotic behaviour of $q_j$ near the singularity is dominated as $j \to \infty$ by the term containing $\Lambda_1$. Setting $Q \sim \Lambda_1 (z-z_s)^3$ near the singular point gives the local behaviour
\begin{equation}\label{e:localansatz}
q_j \sim \frac{\Lambda_1(\tilde{z}-\tilde{z}_s)^3 \Gamma(2j + \gamma)}{[2\pi {\rm i}(\tilde{z} - \tilde{z}_s)]^{2j+\gamma}}\quad \mathrm{as} \quad j\to\infty,\,\tilde{z}\to \tilde{z}_s.
\end{equation}
In order to be consistent with the leading-order asymptotics for $q_0$ from \eqref{e:q0nearzs}, we require that the singularity have strength of 1 when $j = 0$. This gives $\gamma = 4$. An equivalent analysis for the term containing $\Lambda_2$ gives $\gamma = -1$. The large-$j$ asymptotic behaviour of $q_j$ is therefore dominated by the late-order expression in \eqref{e:localansatz}, in which the factorial term has larger argument. 

The constants $\Lambda_1$ and $\Lambda_2$ may be used to determine the constants $K_1$ and $K_2$ in \eqref{e:UK}. We expand this expression for $Q$ in \eqref{e:Uasymp} about the point $\tilde{z} = \tilde{z}_s$ to obtain
\begin{equation}
Q(\tilde{z}) \sim \frac{-\tfrac{\mathrm{i}}{2} K_1 + 3 \pi K_2}{(\tilde{z}-\tilde{z}_s)^2} + \frac{-\tfrac{\mathrm{i}}{2} K_1 + 3 \pi K_2}{6}- \frac{7(-\tfrac{\rm i}{2} K_1 + 3 \pi K_2)}{120 }(\tilde{z}-\tilde{z}_s)^2+ \frac{4 {\rm i}}{5}K_2 (\tilde{z}-\tilde{z}_s)^3 +\ldots
\end{equation}
Comparing this expression with \eqref{e:localU} gives $\Lambda_1 = {4{\rm i}}K_2/5$ and $ \Lambda_2 = -{\rm i}K_1/2 + 3 \pi K_2$. We therefore know that the asymptotic behaviour of $q_j$, described in \eqref{e:ansatz}, as $j \to \infty$ is dominated by factorial-over-power terms with the form
\begin{equation}
q_j \sim \frac{5}{4 {\rm i}}\frac{\Lambda_1 U_2(z) \Gamma(2j + 4)}{ [2\pi {\rm i}(z-z_s)^{2j+4}]} \quad \mathrm{as} \quad j \to \infty.
\end{equation}
In subsequent analysis, we will omit the subscript on and $\Lambda_1$ and $U_2$.

\subsubsection{Calculating $\Lambda$}

To calculate $\Lambda$, we must match the asymptotics of the late-order terms \eqref{e:localansatz} with an inner expansion near the singularity at $\tilde{z} = \tilde{z}_s$. From the expression \eqref{e:series} we see that the series fails to be asymptotic if $q_j(\tilde{z}) = \mathcal{O}(\eps^{-2})$ in the small $\eps$ limit, as the terms are no longer well-ordered. We see from \eqref{e:localansatz} that this corresponds to $\tilde{z} - \tilde{z}_s = \mathcal{O}(\eps)$. Motivated by this scaling, we define a new inner coordinate $\eta$ and inner variable $\hat{q}(\eta)$ such that
\begin{equation}
\tilde{z} - \tilde{z}_s = \eps \eta, \qquad \hat{q}(\eta) = \frac{q(\tilde{z})}{\eps}.
\end{equation}
The rescaled inner equation becomes
\begin{align}
- \eps^2 \omega \hat{q}(\eta) + 2\sum_{m=1}^{\infty} \frac{1}{(2m)!}\frac{d^{2m}\hat{q}(\eta)}{d\eta^{2m}} + |\hat{q}(\eta)|^2\hat{q}(\eta)  &= 0,\label{e:innereq1}\end{align}

Taking only the leading-order asymptotic behaviour of \eqref{e:innereq1} gives the inner system
\begin{align}
2\sum_{m=1}^{\infty} \frac{1}{(2m)!}\frac{d^{2m}\hat{q}(\eta)}{d\eta^{2m}} + |\hat{q}(\eta)|^2\hat{q}(\eta) &= 0.\label{e:innereq3}
\end{align}
The value of $\omega$ does not influence the calculations at this scaling. Note that the inner scaled variable $\eta$ has returned 
the original discrete scaling, and this system is equivalent to
\begin{align}
2[q(\eta+1) -q (\eta) + q(\eta-1)] + |q(\eta)|^2q(\eta)  &= 0.
\end{align}
This is the leading order reduction for the discrete equation \eqref{e:Qn} in the vicinity of singular points. This observation connects the outer scaling in terms of $z$ with the large-$n$ asymptotics of the original discrete problem in terms of $n$.

\begin{figure}[tb]
\centering
 \begin{tikzpicture}
 [x=4*0.85,>=stealth,y=0.5*0.85]
\draw (0,-200) -- (0,-400) -- (101,-400) -- (101,-200) -- cycle;
\draw[->] (101,-200) -- (103,-200) node[right] {\scriptsize{$n$}};
\draw[->] (0,-200) -- (0,-200+8) node[above] {\scriptsize{$\mathrm{Im}(\Lambda_{\mathrm{app}})$}};
\draw [dotted] (0.1,-200) -- (0,-200) node[left]{\scriptsize{$-200$}};
\draw [dotted] (101,-300) -- (0,-300) node[left]{\scriptsize{$-300$}};
\draw [dotted] (0.1,-400) -- (0,-400) node[left]{\scriptsize{$-400$}};
\draw [dotted] (25,-400) -- (25,-200) node[above] {\scriptsize{$25$}};
\draw [dotted] (50,-400) -- (50,-200) node[above] {\scriptsize{$50$}};
\draw [dotted] (75,-400) -- (75,-200) node[above] {\scriptsize{$75$}};
\draw [dotted] (100,-200.1) -- (100,-200) node[above] {\scriptsize{$100$}};
\draw[dashed] (101,-358.513) -- (0,-358.513) node[left] {\scriptsize{-358.513}};
\draw[black] plot[mark=*, only marks,mark size=1pt] file {DNLS-uinner.txt};
 \end{tikzpicture}
\caption{First 100 approximate values of $\Lambda$.}\label{fig:Lambda1}
\end{figure}
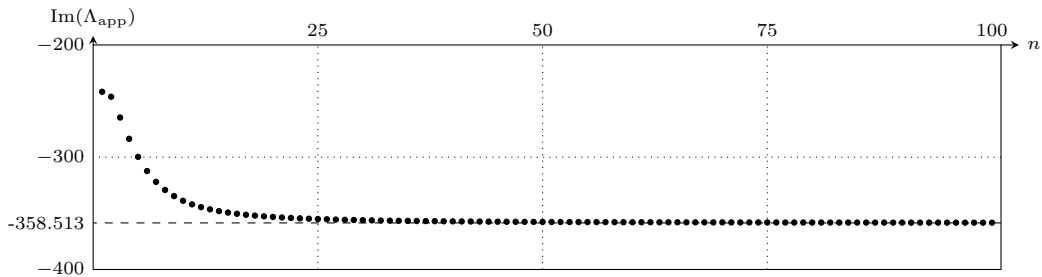

We apply the asymptotic series in the limit that $|\eta| \to \infty$,
\begin{equation}
\hat{q}(\eta) \sim \sum_{j=0}^{\infty}\frac{\hat{q}_j}{\eta^{2j+1}}\quad \mathrm{as} \quad |\eta|\to\infty\label{e:innerseries}
\end{equation}
Based on the local behaviour of leading-order solution \eqref{e:q0nearzs}, we set $q_0 = -{\rm i}\sqrt{2}$. We can then substitute \eqref{e:innerseries} into \eqref{e:innereq3} and balance powers of $\eta$ to obtain a recurrence relation for $q_n$, given by
\begin{align}
2\sum_{j=1}^{n} \binom{2n+2}{2j}\hat{q}_{n-j+1}+ \sum_{j=0}^n\sum_{k=0}^{n-j}\hat{q}_j \hat{q}_k\hat{q}_{n-j-k}&= 0.
\end{align}
These equations may be repeatedly solved for $\hat{q}_n$ up to arbitrary order. To determine the value of $\Lambda$, we match the value of $\hat{q}_n$ for large values of $n$ with the late-order term ansatz \eqref{e:localansatz}. By expressing the inner series \eqref{e:innerseries} in terms of the outer variables, we find that matching $\hat{q}_n$ with the late-order ansatz requires that the asymptotic matching condition
\begin{equation}
\frac{\hat{q}_{n}}{(\tilde{z}-\tilde{z}_s)^{2n+1}} \sim  \frac{\Lambda(\tilde{z}-\tilde{z}_s)^3 \Gamma(2j + 4)}{[2\pi\mathrm{i}(\tilde{z} - \tilde{z}_s)]^{2j+4}}
\end{equation}
be satisfied as $n \to \infty$. By rearranging this, we find that
\begin{equation}\label{e:Lambda1Lim}
\Lambda = \lim_{n\to\infty} \frac{(2\pi{\rm i})^{2n+4} \hat{q}_n}{\Gamma(2j + 4)}.
\end{equation}
To approximate this value, we calculate $\hat{q}_n$ numerically for large values of $n$. These values are then substituted into the expression from \eqref{e:Lambda1Lim} to obtain an approximate value for $\Lambda$, denoted $\Lambda_{\mathrm{app}}$. Figure \ref{fig:Lambda1} shows evidence that $\Lambda_{\mathrm{app}}$ converges as $n \to\infty$. Calculating up to $n = 500$ gives $\Lambda \approx -358.513 {\rm i}$.

\subsection{Stokes Phenomenon}

This procedure is essentially identical to King \& Chapman \cite{king_chapman_2001}; however, that is a nonstandard application of the ideas from Olde Daalhuis et al \cite{Daalhuis}, and we therefore outline the details in Appendix \ref{app:stokes}. 
The general procedure is to truncate the asymptotic series \eqref{e:series} so that the truncation remainder is exponentially small, giving
\begin{equation}\label{e:sertrunc}
q(\tilde{z}) = \sum_{j=0}^{N-1} \eps^{2n}q_n(\tilde{z}) + q_{\mathrm{exp}},
\end{equation}
where $N$ is the truncation value that minimizes the series approximation error, and $q_{\mathrm{exp}}$ is the remainder after truncation, which is exponentially small if $N$ is chosen optimally \cite{BerryHowls1990,Berry1991}; in this case, we find that the correct value is $N = |\chi|/2\eps + \alpha$, where $\alpha \in [0,1)$ is chosen so that $N$ is an integer. We substitute the truncated series \eqref{e:sertrunc} into the governing equation \eqref{e:infode} to obtain an equation for $q_{\mathrm{exp}}$, given in \eqref{e:R1}, that we write in terms of the late-order ansatz \eqref{e:ansatz} to obtain an equation \eqref{e:R12} which we can study using asymptotic methods as $\eps \to 0$.

We apply matched asymptotic expansions to study $q_{\mathrm{exp}}$ in a neighborhood of width $\mathcal{O}(\sqrt{\eps})$ near the Stokes line in order to determine the exponentially small change in the asymptotic contribution that occurs within this neighborhood as the Stokes line is crossed from left to right. We are free to set $q_{\mathrm{exp}}$ to be zero on the left-hand side of the Stokes line, meaning that the change \eqref{e:Sjump2} gives the value of the exponentially small contribution on the right-hand side.

In this analysis we find that for $\tilde{z} > 0$ the exponentially small contribution to the asymptotic behaviour, denoted $q_{\mathrm{exp}}$, satisfies
\begin{equation}\label{e:Rcalc1}
q_{\mathrm{exp}} \sim -\frac{2\pi {\rm i}}{\eps^4}Q e^{-\chi/(\eps\sqrt{\omega})} + \mathrm{c.c.} \quad \mathrm{as} \quad \eps \to 0,
\end{equation}
where we have included the complex conjugate quantity that switches on due to the singularity at $z_{s,-1}$. Simplifying this expression using \eqref{e:chi} and \eqref{e:zs} gives the exponentially small contribution for $\tilde{z}>0$ as
\begin{equation}\label{e:remainder}
q_{\mathrm{exp}} \sim \frac{2\pi}{\eps^4} Q e^{-\pi^2/(\eps\sqrt{\omega})}\sin\left(\frac{2\pi \tilde{z}}{\eps}\right) \quad \mathrm{as} \quad \eps \to 0.
\end{equation}
There is no exponentially small contribution present in the solution for $\tilde{z} < 0$.

\subsection{Site selection}

The oscillations in \eqref{e:remainder} are exponentially small when they switch in amplitude across the Stokes line; however, as $z \to \infty$, the exponentially small oscillations in \eqref{e:remainder} grow as a consequence of to the exponentially growing behaviour of $Q$ described in \eqref{e:Qlimit}. As $\tilde{z} \to \infty$ and $\eps \to 0$, the remainder satisfies the expression
\begin{equation}\label{e:remainder0}
q_{\mathrm{exp}} \sim \frac{2\pi}{\eps^4} \left(\frac{5\Lambda}{4{\rm i}}  e^{\sqrt{\omega}\tilde{z}}\right) e^{-\pi^2/(\eps\sqrt{\omega})}\sin\left(\frac{2\pi \tilde{z}}{\eps}\right).
\end{equation}
We can write this in the original discrete scaling using $\tilde{z} = \eps (n - n_0)$, and using the fact that $\Lambda$ is a negative imaginary number, to obtain
\begin{equation}\label{e:remainder1}
q_{\mathrm{exp}} \sim -\frac{5\pi|\Lambda|}{2\eps^4}   e^{\eps\sqrt{\omega}(n-n_0)} e^{-\pi^2/(\eps\sqrt{\omega})}\sin\left(2\pi (n-n_0)\right)\quad \mathrm{as}\quad n \to \infty,\,\eps \to 0.
\end{equation}
The presence of a slowly growing oscillatory tail in the asymptotic solution indicates that this solution can only exist if $n_0$ is chosen in a way such as to eliminate the oscillations entirely for integer values of $n$. There are only two values in $n_0 \in [0,1)$ that achieve this; $n_0 = 0$, or on-site pinning, and $n_0 = 1/2$, or inter-site pinning.
In this way we retrieve the well-known result
about the existence of solely site-centered and 
inter-site-centered
stationary solutions in the discrete lattice setting~\cite{chriseil,johaub,kev09}.

\section{Stability}
In the previous section we determined that as $\eps \to  0$, if $z > z_0$ then
\begin{equation}
q(\tilde{z}) \sim \sum_{j=0}^{N-1} \eps^{2j} q_j(\tilde{z})  + \frac{2\pi}{\eps^4} Q e^{-\pi^2/(\eps\sqrt{\omega})}\sin\left(\frac{2\pi \tilde{z}}{\eps}\right).
\end{equation}
We will subsequently refer to this steady solution as $q_s(\tilde{z})$. As $\tilde{z} \to \infty$, this asymptotic behavior is 
\begin{equation}
q_s(z) \sim 2\sqrt{2\omega}e^{-\sqrt{\omega}\tilde{z}} + \sum_{j=1}^{N-1} \eps^{2j} q_j(z)  - \frac{2\pi}{\eps^4} \left(\frac{5{\rm i}\Lambda}{4}  \mathrm{e}^{\sqrt{\omega}\tilde{z}}\right)   e^{-\pi^2/(\eps\sqrt{\omega})}\sin\left(\frac{2\pi \tilde{z}}{\eps}\right),
\end{equation}
as ${\tilde{z}}\to \infty$, $\eps \to 0$. We assume that $n_0$, and hence $z_0$, is chosen so that we have either an site-centered or inter-site-centered solution. We introduce perturbations of the form
\begin{equation}
Q = e^{{\rm i}\omega t}\left[q_s + f e^{\lambda t} +g^* e^{\lambda^* t}\right].
\end{equation}
By direct substitution, we obtain
\begin{align}
2\sum_{m=1}^{\infty}\frac{\eps^{2m-2}}{(2m)!}\frac{d^{2m}f}{d\tilde{z}^{2m}} + q_s^2 (2f + g) - \omega f &= -{\rm i}\lambda f,\\
2\sum_{m=1}^{\infty}\frac{\eps^{2m-2}}{(2m)!}\frac{d^{2m}g}{d\tilde{z}^{2m}}  + q_s^2 (2g + f) - \omega g &= {\rm i}\lambda g.
\end{align}
We make the transformation $v = f+g$, $w = f-g$, giving
\begin{align}\label{e:lambdaeq1}
\mathcal{L}_0 v = -{\rm i}\lambda w,\qquad
\mathcal{L}_1 w = {\rm i}\lambda v,
\end{align}
where
\begin{equation}
\mathcal{L}_0 = 2\sum_{m=1}^{\infty}\frac{\eps^{2m-2}}{(2m)!}\frac{d^{2m}}{d{\tilde{z}}^{2m}} + 3q_s^2  - \omega,\qquad 
\mathcal{L}_1 = 2\sum_{m=1}^{\infty}\frac{\eps^{2m-2}}{(2m)!}\frac{d^{2m}}{d{\tilde{z}}^{2m}} + q_s^2  - \omega.
\end{equation}
We expand $v$ and $w$ as power series in the eigenvalue $\lambda$ in the limit $\lambda \to 0$, giving the leading-order equations
\begin{align}
\label{e:v0}\mathcal{L}_0 v_0 = 0,\qquad
\mathcal{L}_1 w_0 &= 0.
\end{align}
We consider two eigenfunctions; one associated with translation and one associated with phase invariance. We can study these by setting
(as is well-known~\cite{Kaup,carretero2025nonlinear})
\begin{equation}
\label{e:translate}v_0  = \frac{dq_s}{{d\tilde{z}}}, \qquad w_0 = 0,
\end{equation}
for the eigenvector associated with translation invariance, and
\begin{equation}
\label{e:phase}v_0 = 0, \qquad w_0 = {\rm i}q_s,
\end{equation}
for the one associated with phase invariance. These choices solve \eqref{e:v0}.

\subsection{First eigenvalue}

\subsubsection{Leading order}

To calculate the eigenvalue associated with translation invariance, we apply \eqref{e:translate} to obtain $w_0 = 0$ and
\begin{equation}
v_0 \sim - \frac{5 |\Lambda|\pi^2}{\eps^5} e^{\sqrt{\omega}\tilde{z}}\mathrm{e}^{-\pi^2/(\eps\sqrt{\omega})} \cos\left(\frac{2\pi\tilde{z}}{\eps}\right)\quad \mathrm{as} \quad \tilde{z} \to \infty,\, \eps \to 0.
\end{equation}
We now balance terms in \eqref{e:lambdaeq1} at $\mathcal{O}(\lambda)$ to obtain
\begin{align}
\mathcal{L}_0 v_1 = 0,\qquad
\mathcal{L}_1 w_1 = -{\rm i}\frac{dq_s}{d\tilde{z}}.
\end{align}
The first equation gives $v_1 = 0$. Using $q_s \sim q_0$ as $\eps \to 0$, we find
\begin{equation}
\label{e:w1b}\frac{d^{2}w_1}{dz^{2}}  + 2 \omega \sech^2(\sqrt{\omega}\tilde{z}) w_1 - \omega w_1 \sim -{\rm i}\sqrt{2\omega}\sech(\sqrt{\omega}\tilde{z})\tanh(\sqrt{\omega}\tilde{z}).
\end{equation}
The homogeneous solutions $W_1$ and $W_2$ are
\begin{equation}
\label{e:W1W2}W_1(z) = \sech(\sqrt{\omega}\tilde{z}), \qquad W_2(z) = \sqrt{\omega}\tilde{z} \sech(\sqrt{\omega}\tilde{z}) + \sinh(\sqrt{\omega}\tilde{z}).
\end{equation}
The general solution is 
\begin{align}
w_1 = -\frac{{\rm i}\sqrt{2}}{2}W_1(\tilde{z})\int_a^{\tilde{z}} &W_2(s)\sech(\sqrt{\omega}s)\tanh(\sqrt{\omega}s)d s\nonumber \\ 
&+ \frac{{\rm i}\sqrt{2}}{2}W_2(\tilde{z})\int_b^{\tilde{z}} W_1(s)\sech(\sqrt{\omega}s)\tanh(\sqrt{\omega}s) ds.\label{e:w1c}
\end{align}
where we are free to choose $a$ and $b$. These integrals can be evaluated exactly using \eqref{e:W1W2}. The function $W_2(z) \to \infty$ as $z \to \pm\infty$. We therefore select $b$ so that the second integral vanishes in this limit, or $b = -\infty$. The resulting expression is
 \begin{equation}
 w_1= -\frac{{\rm i}\sqrt{2\omega}}{2} \big(\tilde{z}\sech(\sqrt{\omega}\tilde{z}) + \alpha W_1(\tilde{z})\big),
 \end{equation}
 where $\alpha$ is a constant determined by the choice of $a$ in \eqref{e:w1c}. As $W_1(z)$ decays in both directions, we do not obtain a condition on the choice of $\alpha$. The resulting expression for $w_1(\tilde{z})$ decays as $\tilde{z} \to \pm\infty$.
 
 Balancing terms in \eqref{e:lambdaeq1} at $\mathcal{O}(\lambda^2)$ gives
 \begin{align}
\label{e:v2}\mathcal{L}_0 v_2 = -{\rm i}w_1,\qquad
\mathcal{L}_1 w_2 = 0.
\end{align}
We see that $w_2 = 0$, and $v_2$ satisfies
\begin{equation}
\frac{d^{2}v_2}{d\tilde{z}^{2}} + (3u_s^2 -\omega)  v_2 = -\frac{\sqrt{2\omega}}{2}(\tilde{z} + \alpha)\sech(\sqrt{\omega}\tilde{z}).
\end{equation}
The homogeneous solutions to this equation were given previously in \eqref{e:U1}--\eqref{e:U2}. For consistency in notation within this section, we define
\begin{align}
V_1(\tilde{z}) &= \tfrac{1}{2} \tanh(\sqrt{\omega}\tilde{z})\sech(\sqrt{\omega}\tilde{z}),\label{e:V1}\\
V_2(\tilde{z}) &= \sech(\sqrt{\omega}\tilde{z})\,\left(6\sqrt{\omega}\tilde{z}\tanh(\sqrt{\omega}\tilde{z}) + \cosh(2\sqrt{\omega}\tilde{z})-5\right).\label{e:V2}
\end{align}
The general solution is 
\begin{align}
v_2 = \frac{1}{2\sqrt{2}}V_1(\tilde{z})\int_a^{\tilde{z}} &V_2(s)(s + \alpha)\sech(\sqrt{\omega}s)ds\nonumber \\ 
&- \frac{1}{2\sqrt{2}}V_2({\tilde{z}})\int_b^{\tilde{z}} V_1(s)(s + \alpha)\sech(\sqrt{\omega}s) ds,\label{e:v1c}
\end{align}
where we are free to choose $a$ and $b$. The first term decays as $V_1({\tilde{z}})$, so we do not obtain any conditions on $a$. The second term grows as ${\tilde{z}} \to \pm\infty$. We can select $b$ to suppress growth on one side, but not both. We choose $b = -\infty$, which ensures that the second integral vanishes as ${\tilde{z}} \to -\infty$. As ${\tilde{z}} \to \infty$, we have
\begin{align}\label{e:lambda2}
v_2 \sim -\frac{1}{2\sqrt{2}}V_2({\tilde{z}})\int_{-\infty}^{\tilde{z}} V_1(s)(s + \alpha)\sech(\sqrt{\omega}s) ds = -\frac{V_2({\tilde{z}})}{4\sqrt{2}} \sim -\frac{e^{{\sqrt{\omega}\tilde{z}}}}{4\sqrt{2}}.
\end{align}
We are now able to predict the behaviour of the tail as $\tilde{z} \to \infty$. On lattice points, we have 
\begin{equation}
v \sim v_0 + \lambda^2 v_2 \sim - \frac{5 |\Lambda|\pi^2}{\eps^5} e^{\sqrt{\omega}\tilde{z}} e^{-\pi^2/(\eps\sqrt{\omega})} \cos\left(\frac{2\pi\tilde{z}}{\eps}\right) - \frac{\lambda^2 e^{\sqrt{\omega}\tilde{z}}}{4\sqrt{2}}\quad \mathrm{as} \quad \tilde{z} \to \infty,\, \eps \to 0.
\end{equation} 
Both terms grow large in the limit $z \to \infty$, and can only be avoided if the terms cancel precisely. This gives the condition that 
\begin{equation}
    v_0 \sim -\lambda^2 v_2 \quad \mathrm{as} \quad \eps \to 0.
\end{equation}
Solving this expression for $\lambda$ and writing the result in terms of the original discrete variable $n$ gives
\begin{align}
\lambda^2 \sim -\frac{40\sqrt{2}|\Lambda|\pi^2}{\eps^{5}} e^{-\pi^2/(\eps\sqrt{\omega})}\cos\left({2\pi(n-n_0)}\right)\quad \mathrm{as} \quad \eps \to 0.
\end{align}
For on-site pinning  $n_0$ is an integer, so $\lambda^2 < 0$. The eigenvalues are imaginary, and the on-site standing wave is stable. For inter-site pinning $n_0$ is a half-integer, so there is a positive real value for $\lambda$ and the corresponding waveform is unstable. In both cases, the eigenvalue has magnitude as $\eps \to 0$
\begin{align}|\lambda|\sim \frac{2^{7/4}5^{1/2}\pi|\Lambda|^{1/2} e^{-\pi^2/(2\eps\sqrt{\omega})}}{\eps^{5/2}}\approx \frac{316.355 e^{-\pi^2/(2\eps\sqrt{\omega})}}{\eps^{5/2}} .
\label{original}
\end{align}
To validate this prediction numerically, we will require the first correction term to $\lambda$ in the limit $\eps \to 0$. 
It is also worthwhile to note that this prediction shares both the
exponential dependence and the prefactor 
$\varepsilon$-scaling of the 
work of~\cite{Todd_Kapitula_2001}, yet has a different numerical
prefactor than the $116.2$ of the latter work.

\subsubsection{First correction term}

To determine the correction to the eigenvalue, we must match the asymptotic solution for $q$ with two terms of an inner expansion in the neighborhood of the singularity at $\tilde{z} = \tilde{z}_s$. To do this, we apply the trick introduced in \cite{king_chapman_2001} and find the first correction to the pole locations in $q(\tilde{z})$.

Combining the local expansions from \eqref{e:q0nearzs} and \eqref{e:q1nearzs} near $\tilde{z} = \tilde{z}_s$ gives
\begin{equation}
q \sim -\frac{{\rm i}\sqrt{2}}{\tilde{z}-\tilde{z}_s} + \frac{{\rm i}\omega}3{\sqrt{2}}(\tilde{z}-\tilde{z}_s) + \eps^2\left(\frac{{\rm i}\sqrt{2}}{3(\tilde{z}-\tilde{z}_s)^3} + \frac{\pi\sqrt{\omega}}{24\sqrt{2}(\tilde{z}-\tilde{z}_s)^2}\right) \quad \mathrm{as} \quad \eps \to 0,\, \tilde{z} \to \tilde{z}_s.
\end{equation}
Expressing this in terms of the inner variables $\eps \eta = \tilde{z}-\tilde{z}_s$ and $\hat{q}(\eta) = \eps q(\tilde{z})$, we find that
\begin{equation}
\hat{q}(\eta) \sim \left(-\frac{{\rm i}\sqrt{2}}{\eta} + \frac{{\rm i}\sqrt{2}}{3\eta^2}\right) + \frac{\eps\pi\sqrt{\omega}}{24\sqrt{2}\eta} + \mathcal{O}(\eps^2)
\end{equation}
as $\eps \to 0$. We instead define the alternative inner variable $\eps \hat{\eta} = \tilde{z} - \tilde{z}_{s} - \pi {\rm i}\sqrt{\omega}\eps^2/48$.This inner variable is chosen to precisely cancel the $\mathcal{O}(\eps)$ correction term, leaving
\begin{equation}
\hat{q}(\eta) \sim \left(-\frac{{\rm i}\sqrt{2}}{\hat{\eta}} + \frac{{\rm i}\sqrt{2}}{3\hat{\eta}^2}\right) + \mathcal{O}(\eps^2)
\end{equation}
as $\eps \to 0$, which is accurate to the required two terms. This tells us that the true pole location has an $\mathcal{O}(\eps^2)$ correction, and we set $\tilde{z}_{s,0} = {\rm i}\pi/2\sqrt{\omega}+ {\rm i}\pi\sqrt{\omega}\eps^2/48 $. This adjusts the value of $\chi = 2\pi{\rm i}(\tilde{z}-\tilde{z}_s)$ in \eqref{e:Rcalc}, 
which in turn produces the adjusted exponentially-small remainder as $\eps \to 0$ and $\tilde{z} \to \infty$
\begin{align}\label{e:remainder1st}
R &\sim -\frac{5\pi|\Lambda|}{2\eps^4} e^{\sqrt{\omega}\tilde{z}} e^{-\pi^2/(\eps\sqrt{\omega})} e^{- \eps\pi^2\sqrt{\omega}/48}\sin\left(\frac{2\pi \tilde{z}}{\eps}\right)\\
&\sim   -\frac{5\pi|\Lambda|}{2\eps^4} e^{\sqrt{\omega}\tilde{z}} e^{-\pi^2/(\eps\sqrt{\omega})}\left(1 - \frac{\eps\pi^2\sqrt{\omega}}{24}\right)\sin\left(\frac{2\pi \tilde{z}}{\eps}\right)
\end{align}
This allows us to determine the first correction in $\eps$ to the eigenvalue $\lambda$. We find that
\begin{align}|\lambda|&\sim \frac{2^{7/4}5^{1/2}\pi|\Lambda|^{1/2} e^{-\pi^2/(2\eps\sqrt{\omega})}}{\eps^{5/2}} \left(1 - \frac{\eps\pi^2\sqrt{\omega}}{24}\right)^{1/2}\\
&\sim \frac{2^{7/4}5^{1/2}\pi|\Lambda|^{1/2} e^{-\pi^2/(2\eps\sqrt{\omega})}}{\eps^{5/2}} \left(1 - \frac{\eps\pi^2\sqrt{\omega}}{48}\right)\quad \mathrm{as} \quad \eps \to 0.
\end{align}
Evaluating this expression gives the numerical approximation
\begin{equation}
 |\lambda|   \approx 
316.355 \left(1 - \frac{\eps\pi^2\sqrt{\omega}}{48}\right)\frac{\mathrm{e}^{-\pi^2/(2\eps\sqrt{\omega})}}{\eps^{5/2}}.
\label{correction}
\end{equation}

\subsubsection{Numerics}

\begin{figure}
\centering
  \includegraphics[width=0.45 \columnwidth]{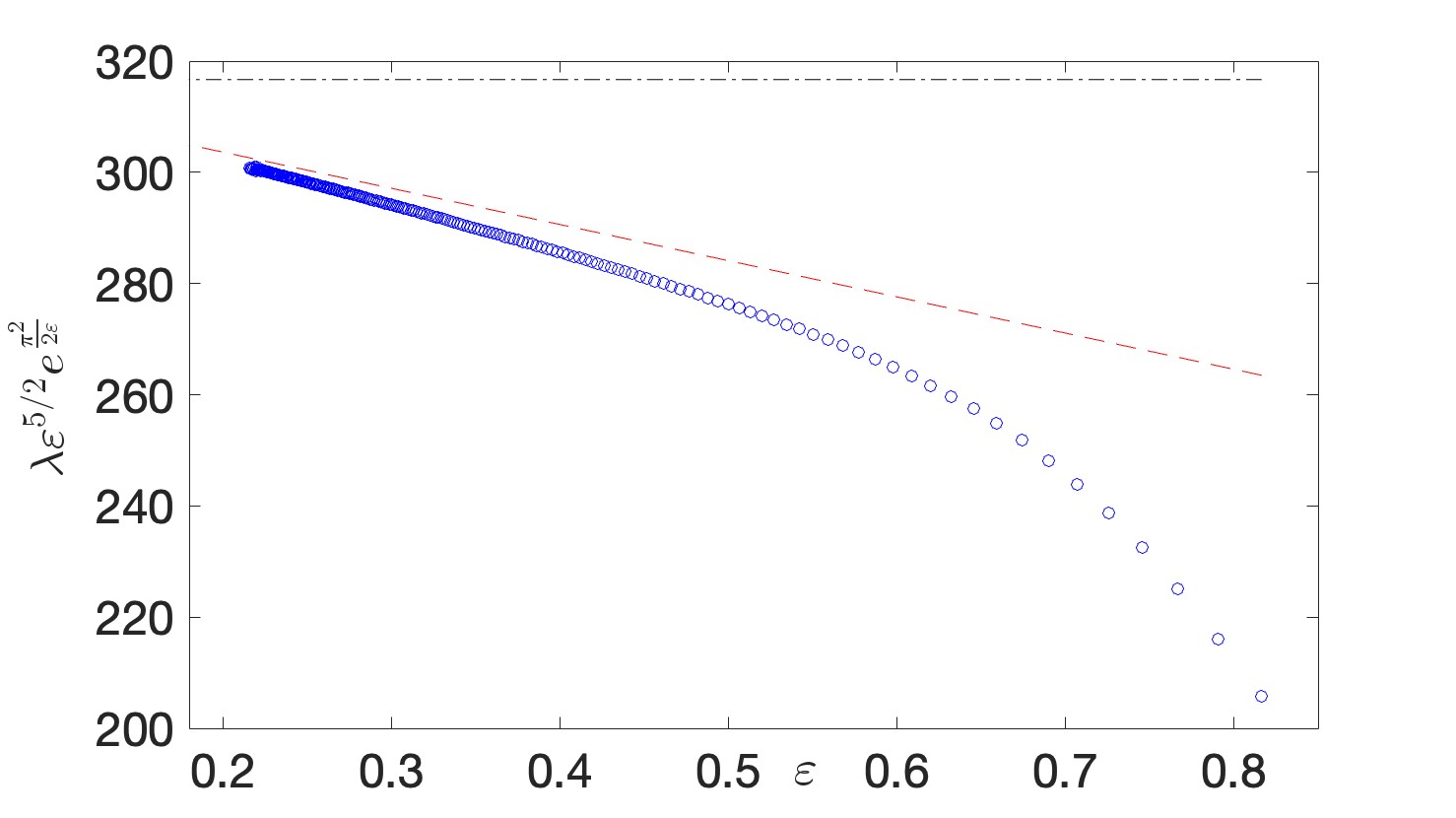}
  \includegraphics[width=0.45 \columnwidth]{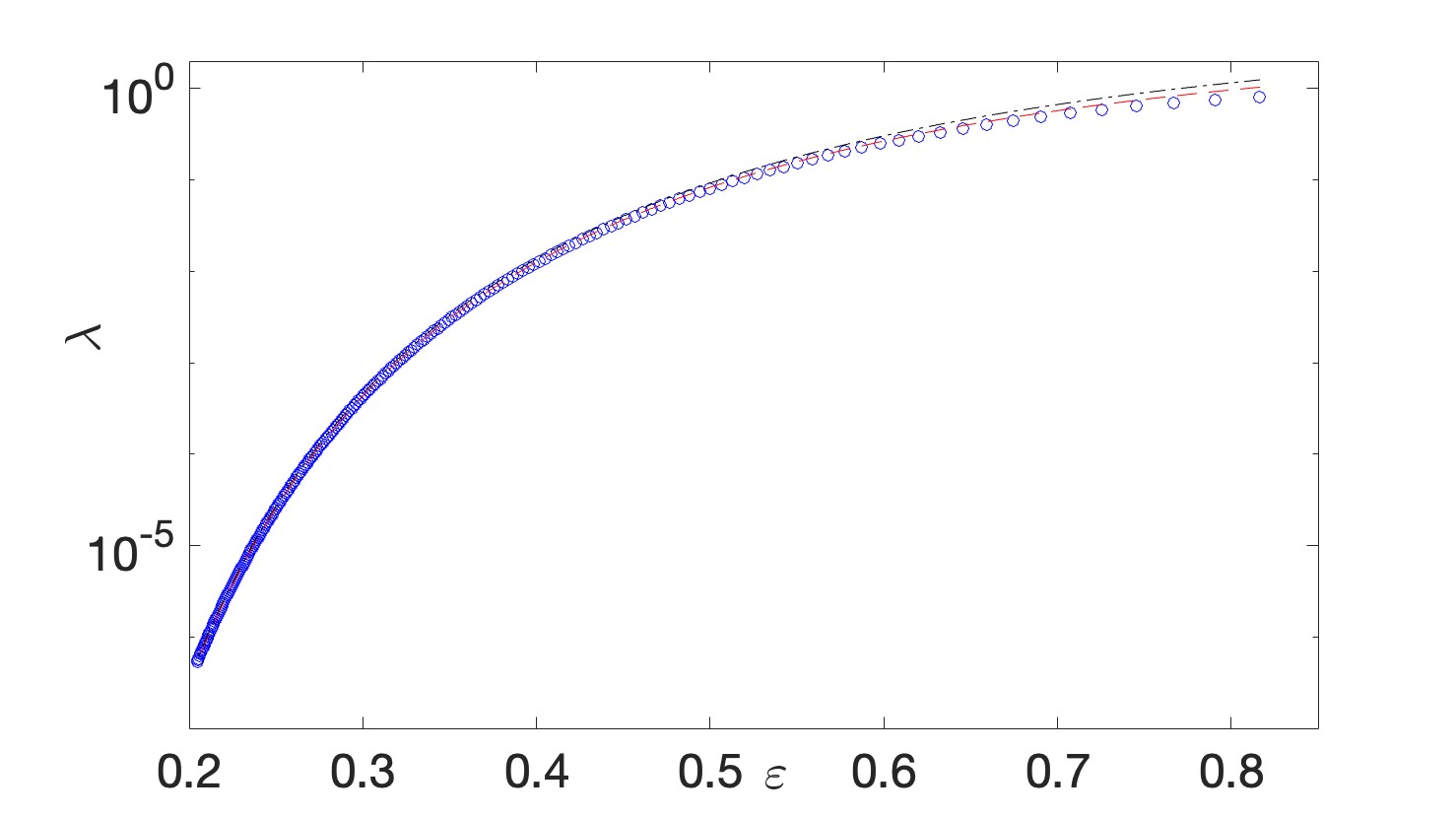}
    \caption{In the left panel we show the asymptotic predictions of
    our analysis herein, namely Eq.~(\ref{original}) by a black
    dash-dotted line, and its leading order correction
    of Eq.~(\ref{correction}) by a red dashed line. These eigenvalue
    predictions (as a function of the lattice spacing $\eps$)
    have been
    multiplied by $\eps^{5/2} \mathrm{e}^{\pi^2/(2\eps\sqrt{\omega})}$, so
    that the former one amounts to a constant and the latter to a
    straight line. These are compared to the correspondingly plotted
    eigenvalues associated with the site-centered solitary wave,
    plotted by blue circles. To illustrate how stringent of a test
    the left panel represents and showcase how close these 3 curves
    are as concerns the original eigenvalue itself, we can observe
    the same predictions/findings for $\lambda$ as a function of
    the coupling $\eps$ in the right panel.}
    \label{lfig3}
\end{figure}

In order to examine the validity of the original
prediction of Eq.~(\ref{original}) and the associated first
correction term of Eq.~(\ref{correction}), we have performed
computations of the relevant eigenvalues for the case of
the site-centered solitary waves and their associated stability.
Such computations are admittedly not new, as they have earlier
appeared in~\cite{johaub,Todd_Kapitula_2001,kev09}, however
the distinguishing feature here is that we push further towards
the relevant limit than was previously considered and, in addition,
we plot them in a way that clearly brings forth the agreement
with our theoretical predictions. Indeed, looking already
at the right panel of Fig.~\ref{lfig3}, the agreement between
the original prediction of Eq.~(\ref{original}) (black dash-dotted
line) and the full numerical results seems very good. This
agreement seems to be even further improved by the presence of
the first correction term (red dashed line). 
It is important to appreciate here that
the variation of the relevant eigenvalue
occurs over more than 6 orders of magnitude
for the interval of spacings $\varepsilon$ shown.

However, the true accuracy of our findings is much more clearly
underscored by the far more demanding test of multiplying
the eigenvalue findings by $\eps^{5/2} \mathrm{e}^{\pi^2/(2\eps\sqrt{\omega})}$. This rescaling not only strips the
theoretical prediction of its dominant functional dependence,
clearly distinguishing between the original form
of Eq.~(\ref{original}) (which is now just a constant)
and of Eq.~(\ref{correction}) (which is now just a straight line).
It also imposes a multiplication by a {\it growing exponential}
which would be extremely sensitive towards a potential error
in the relevant exponent. This clearly confirms not only the accuracy
of the exponential dependence and of the power law prefactor;
it also showcases the systematic approach to our asymptotic
prediction in a quite definitive manner.
It is relevant to note here that we have use a diverse palette
of tools for obtaining these eigenvalues accurately to about $10^{-7}$.
Seeking a further probing of the asymptotics requires tools
that are considerably more refined at the level of the eigenvalue solver which, while
not impossible, is outside the scope of the present work.

\subsection{Second eigenvalue}

We may repeat the previous procedure in order to determine the second eigenvalue, using the  perturbation
\begin{equation}
v_0 = 0, \qquad w_0 = {\rm i}u_s.
\end{equation}
Following the analysis from before, $w_1 = 0$ and the equation for $v_1$ at leading order as $\lambda, \eps \to 0$ is
\begin{equation}
\frac{d^2 v_1}{d\tilde{z}^2} + (3 u_s^2  -\omega) v_1 = u_0.
\end{equation}
The homogeneous solutions are identical to \eqref{e:V1}--\eqref{e:V2}. 
The general solution is given by
\begin{equation}
    w_2 = -\frac{V_1(\tilde{z})}{\sqrt{2}}\int_a^{\tilde{z}}V_2(s)\sech(s)\,ds + \frac{V_2(\tilde{z})}{\sqrt{2}}\int_b^{\tilde{z}}V_1(s)\sech(s)\,ds.
\end{equation}
We select $b = -\infty$ to suppress growth $V_2$ as $\tilde{z} \to -\infty$, giving
\begin{equation}
v_1 = \frac{\sech(\sqrt{\omega}{\tilde{z}})}{\sqrt{2\omega}}\left(1- \sqrt{\omega}(\tilde{z}+\alpha)\tanh(\sqrt{\omega}\tilde{z}) \right),
\end{equation}
where $\alpha$ is a constant that depends on the choice of $a$. Balancing terms at $\mathcal{O}(\lambda^2)$, we find $v_2 = 0$ and
\begin{equation}
\frac{d^2 w_2}{dz^2} +  (u_s^2  - \omega)w_2 = - v_1.
\end{equation}
The homogeneous solutions are given in \eqref{e:W1W2}. The general solution is
\begin{align}\nonumber
    w_2 = \frac{\rm i}{2\sqrt{2\omega}}W_1(\tilde{z})&\int_a^{\tilde{z}}W_2(s)\sech(\sqrt{\omega}s)(1 - \sqrt{\omega}(s+\alpha)\tanh(\sqrt{\omega}s))\, ds\\
    &-\frac{\rm i}{2\sqrt{2\omega}}W_2(\tilde{z})\int_b^{\tilde{z}}W_1(s)\sech(\sqrt{\omega}s)(1 - \sqrt{\omega}(s+\alpha)\tanh(\sqrt{\omega}s))\, ds
\end{align}
As before, we suppress the growing $W_1$ term as $\tilde{z}\to-\infty$ by setting $b = -\infty$. As $\tilde{z} \to \infty$, the first integral vanishes for any choice of $a$. This leaves
\begin{equation}
w_2 \sim -\frac{\rm i}{2\sqrt{2\omega}}W_2(\tilde{z})\int_{-\infty}^{\infty}W_1(s)\sech(\sqrt{\omega}s)(1 - \sqrt{\omega}(s+\alpha)\tanh(\sqrt{\omega}s))\, ds = -\frac{{\rm i}W_2(\tilde{z})}{2\sqrt{2}} \sim -\frac{{\rm i} e^{\sqrt{\omega}\tilde{z}}}{4\sqrt{2}}.
\end{equation}
We now calculate
\begin{equation}\label{e:lambda2phase}
w \sim w_0 + \lambda^2 w_2 \sim   - \frac{5{\rm i}|\Lambda|\pi}{2\eps^4} e^{\sqrt{\omega}\tilde{z}} e^{-\pi^2/(\eps\sqrt{\omega})} \sin\left(\frac{2\pi\tilde{z}}{\eps}\right) - \frac{{\rm i}\lambda^2 e^{\sqrt{\omega}\tilde{z}}}{4\sqrt{2}}\quad \mathrm{as} \quad \tilde{z} \to \infty,\, \eps \to 0.
\end{equation} 
As this expression grows exponentially as $\tilde{z} \to \infty$, we need to determine $\lambda$ such that $w_0 + \lambda^2 w_2 = 0$. The expression in \eqref{e:lambda2phase} is similar to \eqref{e:lambda2}; however, there is an important distinction. In this case, the first term  contains
\begin{equation}
\sin\left(\frac{2\pi\tilde{z}}{\eps}\right) = \sin(2\pi(n-n_0)).
\end{equation}
The value of $n_0$ can be an integer (on-site pinning) or a half-integer (inter-site pinning). In both cases, this term is equal to zero. This means that the only value of $\lambda$ that solves $w_0 + \lambda^2 w_2 = 0$ is $\lambda = 0$. This is naturally what is expected in this case: in 
particular (and contrary to what is the case for the translational 
invariance leading to the eigenvalue computed in section IV.A),
here the eigenvector pertains to a symmetry that is preserved
in the presence of discreteness. Hence, we expect the corresponding
eigenvalue to indeed ``stay put'' at $\lambda^2=0$, in line
with our numerical observations.

\section{Conclusions \& Outlook}

In the present work we have revisited the time-honored
problem of the breaking of translational invariance and
its existence and spectral stability implications
in nonlinear lattice dynamical equations. We have used
as a prototype for our analysis the discrete nonlinear
Schr{\"o}dinger equation, although we have alluded throughout
our presentation to the generality of our considerations,
for instance in the large class of models that involves
discretizations of the Laplacian operator, such as,
for instance, discrete Klein-Gordon, as well as other
discrete NLS variants. Our exponential asymptotics
formulation has enabled us to infer the ability of
the lattice to center the solution either on-site or
inter-site at the level of the existence problem.
The subsequent consideration of these states in the
stability analysis has prompted us to conclude the
stability of the former and instability of the latter,
but importantly to also offer definitive information
on not only the exponential dependence (on the lattice
spacing), but also its power law form multiplying it.
Indeed, our work went far further than these features
that were previously known, e.g., to~\cite{Todd_Kapitula_2001}.
More specifically the prefactor of this composite
(exponential multiplying power law) dependence was
accurately identified and the demanding leading
order correction to this asymptotic form was also 
inferred and confirmed through numerical computations
extending over several orders of magnitude of eigenvalue data.

Naturally, these considerations and the accuracy of the
exponential asymptotics in capturing such eigenvalues
raise numerous further questions and generate various
possibilities for future studies. As some of the most
canonical ones among them, we note the following.
Here, we have restricted considerations to the standard
Laplacian term involving nearest neighbor interactions.
However, in optical waveguide array applications the
consideration of next-nearest-neighbor lattices involving
the so-called zigzag waveguide arrays has not only
been theoretically proposed~\cite{Efremidis2002}; it has
been experimentally implemented, e.g., in the work of~\cite{Szameit:09}.
This naturally poses the question of the potential interplay
(and indeed perhaps even competition) between the role of
the nearest- and next-nearest-neighbor interactions.
While this is technically considerably more involved due to the presence of additional exponentially small asymptotic contributions, it
is well within the purview of our theoretical methodology.
In an additional vein, works such as~\cite{Todd_Kapitula_2001}
and before it also such as~\cite{Peypel,KapKevJon} have
studied bifurcations of internal modes of solitary waves
and have illustrated that these combine (pure) power-law
effects in the lattice spacing~\cite{Peypel} with
(modulated) exponential ones. It remains an important
open question whether variants of the method proposed
herein can accurately, not just at a qualitative level,
but also at a quantitative one as herein, capture the
relevant dependence. Finally, it is natural to recall
that integrable variants of the DNLS model remain of
considerable interest for both computational and
perturbative analysis purposes. The so-called Salerno-model
homotopic continuation between the two limits (DNLS and
Ablowitz-Ladik)~\cite{salerno1992quantum} remains a popular 
setting in its own right. Thus, an understanding of the
translational mode behavior homotopically between the
two limits in a quantitative fashion is naturally of
interest. Of course, extensions of the methodology to 
DNLS models in higher dimensions~\cite{kev09}
(where excitation thresholds~\cite{Kladko,Weinstein_1999}
and associated stability changes arise) would naturally
be of interest as well. Such studies are presently
under consideration and will be reported in future publications.

{\it Acknowledgements.}
This material is based upon work supported by the U.S. National Science Foundation under the awards PHY-2110030, PHY-2408988 and DMS-2204702 (PGK), and Australian Research Council DP190101190 and DP240101666 (CJL).
Additionally, PGK gratefully acknowledges past
discussions on this topic  with
C.K.R.T. Jones and T. Kapitula (from 1999 to 2001)
and with I. Mylonas, as well as H. Susanto and
R. Kusdiantara (from 2016 to 2018) that
motivated the present work. 

\bibliographystyle{plain}
\bibliography{reference2}

\appendix

\section{Stokes switching calculations}\label{app:stokes}

Truncating the series \eqref{e:series} after $N$ terms gives
\begin{equation}
q(\tilde{z}) = \sum_{j=0}^{N-1} \eps^{2n}q_n(\tilde{z}) + R(\tilde{z}).
\end{equation}
We follow the heuristic described by Boyd \cite{Boyd1999} and choose the value of $N$ that minimizes \eqref{e:ansatz} over $j$. This gives $N = \tfrac{|\chi|}{2\eps} + \alpha$, where $\alpha \in [0,1)$ is chosen so that $N \in \mathbb{Z}$. $R(\tilde{z})$ is the remainder after truncation; if $N$ is chosen optimally, then this quantity is exponentially small as $\eps \to 0$. After simplification, we have 
\begin{align}\label{e:R1}
&2\sum_{m=1}^{\infty}\frac{\eps^{2m-2}}{(2m)!}\frac{d^{2m}R}{d\tilde{z}^{2m}} + (3 q_0^2 - \omega) R  \sim -2\sum_{p=N}^{\infty}\eps^{2p}\sum_{m=p-N+2}^{p+1}\frac{1}{(2m)!}\frac{d^{2m}q_{p-m+1}}{d\tilde{z}^{2m}}.
\end{align}
Applying a WKB analysis to the homogeneous version of \eqref{e:R1} gives $R \sim k Q\mathrm{e}^{-\chi/\eps}$ where $k$ is an arbitrary constant. Using variation of parameters therefore suggests 
\begin{equation}
R \sim \mathcal{S} Q\mathrm{e}^{-\chi/\eps} \quad \mathrm{to}\quad \eps \to 0,
\end{equation}
where $\mathcal{S}$ behaves as a constant everywhere except in the neighborhood of a Stokes line, where it changes rapidly in value. Substituting this expression into \eqref{e:R1} gives after simplification
\begin{align}
2\left[\sum_{m=1}^{\infty}\binom{2m}{2}\frac{(\chi')^{2m-2}}{(2m)!}\right]\frac{d^2}{d\tilde{z}^2}\left(\mathcal{S} Q\right)\mathrm{e}^{-\chi/\eps}+ (3 u_0^2 - \omega) \mathcal{S}Q\mathrm{e}^{-\chi/\eps} \sim -2\sum_{p=N}^{\infty}\eps^{2p}\sum_{m=p-N+2}^{p+1}\frac{1}{(2m)!}\frac{d^{2m}q_{p-m+1}}{d\tilde{z}^{2m}}.
\end{align}
We substitute in $\chi' = 2\pi\mathrm{i}$ and the late-order ansatz \eqref{e:ansatz} for $u_j$ to give
\begin{equation}\label{e:R12}
\left[\frac{d^2}{d{\tilde{z}}^2}\left(\mathcal{S} Q\right)+(3 u_0^2 - 1) \mathcal{S} Q \right]\mathrm{e}^{-\chi/\eps} \sim -2\sum_{p=N}^{\infty}\eps^{2p}\sum_{m=p-N+2}^{p+1}\frac{(\chi')^{2m}Q\Gamma(2p+\gamma+2)}{(2m)!\chi^{2p+\gamma+2}},
\end{equation}
where the omitted terms are smaller as $\eps \to 0$ and $N \to \infty$. 

We can extend the second summation term to infinity as before, as $p \geq N$ which is asymptotically large as $\eps \to 0$. Hence we write
\begin{equation}
\left[\frac{d^2}{d{\tilde{z}}^2}\left(\mathcal{S} Q\right)+(3 q_0^2 - \omega) \mathcal{S} Q \right]\mathrm{e}^{-\chi/\eps}\sim -2\sum_{p=N}^{\infty}\eps^{2p}\sum_{m=p-N+2}^{\infty}\frac{(\chi')^{2m}Q\Gamma(2p+\gamma+2)}{(2m)!\chi^{2p+\gamma+2}},
\end{equation}
To make the scale of $p$ apparent, we write $p = N + q$. Swapping the order of summation and writing $\chi = r\mathrm{e}^{\mathrm{i}\theta}$ and $N = {r}/{2\eps} + \alpha$ gives
\begin{equation}
\left[\frac{d^2}{d{\tilde{z}}^2}\left(\mathcal{S} Q\right)+(3 q_0^2 - \omega) \mathcal{S} Q \right]\mathrm{e}^{-\chi/\eps} \sim -2\sum_{m=2}^{\infty}\sum_{q=0}^{m-2}\eps^{r/\eps+2q + 2\alpha}\sum_{m=p-N+2}^{\infty}\frac{(2\pi\mathrm{i})^{2m}Q\Gamma(r/\eps+2q+\gamma+2+2\alpha)}{(2m)!\chi^{r/\eps+2q+\gamma+2+2\alpha}}.
\end{equation}
Applying Stirling's formula gives
\begin{equation}
\left[\frac{d^2}{d{\tilde{z}}^2}\left(\mathcal{S} Q\right)+(3 q_0^2 - \omega) \mathcal{S} Q \right]\mathrm{e}^{-\chi/\eps} \sim -2\sum_{m=2}^{\infty}\sum_{q=0}^{m-2}\frac{(2\pi\mathrm{i})^{2m}Q\sqrt{2\pi}\mathrm{e}^{-r/\eps}}{(2m)!\eps^{\gamma+3/2}r^{1/2}}\mathrm{e}^{-\mathrm{i}\theta(r/\eps + 2q + \gamma+2+2\alpha)}.
\end{equation}
We can evalute these sums exactly to obtain
\begin{equation}
\frac{d^2}{d{\tilde{z}}^2}\left(\mathcal{S} Q\right)+ (3q_0^2 - \omega) \mathcal{S} Q  \sim \frac{2 \mathrm{e}^{-r/\eps - \mathrm{i}\theta(r/\eps + \gamma+2\alpha)}\sqrt{2\pi}Q f(\theta)}{\eps^{\gamma+3/2}r^{1/2}}\mathrm{e}^{\chi/\eps},
\end{equation}
where
\begin{equation}
f(\theta) =\frac{\cos(2\pi\mathrm{e}^{-\mathrm{i}\theta})-1}{1-\mathrm{e}^{-2\mathrm{i}\theta}}.
\end{equation}
Note that $f(\theta) \sim -\mathrm{i}\pi^2\theta$ as $\theta \to 0$. 
We consider variation in the $\theta$ direction, and set
\begin{equation}
\frac{d}{d{\tilde{z}}} = \chi'\frac{d}{d\chi}, \qquad \frac{d}{d\chi} = -\frac{\mathrm{i}}{r}\mathrm{e}^{-\mathrm{i}\theta}\frac{d}{d\theta}.
\end{equation}
Some algebra gives
\begin{equation}
\frac{d^2}{d{\tilde{z}}^2}\left(\mathcal{S} Q\right) = -\frac{(\chi')^2}{r^2}\mathrm{e}^{-\mathrm{i}\theta}\frac{d}{d\theta}\left(\mathrm{e}^{-\mathrm{i}\theta}\frac{d}{d\theta}\left(\mathcal{S} Q\right)\right).
\end{equation}
The rapid variation across the Stokes line in the $\theta$ direction is captured by the change in $\mathcal{S}$; $Q$ does not change rapidly in this direction. This means that $(\mathcal{S}Q)_{\theta}\sim Q\mathcal{S}_{\theta}$ as $\eps \to 0$. Hence
\begin{equation}
\mathrm{e}^{-\mathrm{i}\theta}\frac{d}{d\theta}\left(\mathrm{e}^{-\mathrm{i}\theta}\frac{d}{d\theta}\left(\mathcal{S}\right)\right) \sim -\frac{r^2}{(2\pi\mathrm{i})^2}\frac{2 \sqrt{2\pi}f(\theta)}{\eps^{\gamma+3/2}r^{1/2}}\exp\left(-\frac{r}{\eps} - \mathrm{i}\theta\left(\frac{r}{\eps} + \gamma+2\alpha\right) + \frac{r}{\eps}\mathrm{e}^{\mathrm{i}\theta}\right).
\end{equation}
To determine rapid variation in a neighborhood of the Stokes line, we set $\theta = \eps^{1/2}\vartheta$. In this region, 
\begin{equation}
\frac{1}{\eps}\frac{d^2\mathcal{S}}{d\theta^2} \sim -\frac{r^2}{(2\pi\mathrm{i})^2}\frac{2 \sqrt{2\pi}(-\mathrm{i}\pi^2\eps^{1/2}\vartheta)}{\eps^{\gamma+3/2}r^{1/2}}\exp\left(-\frac{r}{\eps} - \mathrm{i}\eps^{1/2}\vartheta\left(\frac{r}{\eps} + \gamma+2\alpha\right) + \frac{r}{\eps}\mathrm{e}^{\mathrm{i}\eps^{1/2}\vartheta}\right),
\end{equation}
where we have used that $f(\eps^{1/2}\vartheta) \sim -\mathrm{i}\pi^2\eps^{1/2}\vartheta$ as $\eps \to 0$. Simplification and integrating twice gives
\begin{equation}\label{e:Sjump}
\frac{d\mathcal{S}}{d\theta} \sim -\frac{{\rm i}}{\eps^{\gamma}}\sqrt{\frac{\pi r}{2}}\int_{-\infty}^{\vartheta}\mathrm{e}^{-rs^2}ds.
\end{equation}
We denote the jump in a quantity across the Stokes line from $\mathrm{Im}(\chi)< 0$ to $\mathrm{Im}(\chi)>0$ as $[\,\cdot\,]_-^+$, so \eqref{e:Sjump} gives
\begin{equation}\label{e:Sjump2}
\left[\mathcal{S}\right]_-^+ \sim \frac{\pi \mathrm{i}}{\eps^\gamma} \quad \mathrm{as} \quad \eps \to 0.
\end{equation}
We will impose the condition that the remainder is present on the right-hand side of the Stokes line, corresponding to $\tilde{z}>0$. Hence for ${\tilde{z}} > 0$ we have 
\begin{equation}\label{e:Rcalc}
R \sim -\frac{2\pi \mathrm{i}}{\eps^{\gamma}}Q\mathrm{e}^{-\chi/\eps}.
\end{equation}

\end{document}